\documentclass[twocolumn, prl, showpacs,superscriptaddress,floatfix]{revtex4}
\usepackage{graphicx}
\usepackage{dcolumn}
\usepackage{bm}
\usepackage{amsmath}

\begin{document}

\title{Ground State Energy for Fermions in a 1D Harmonic Trap\\
with Delta Function Interaction }

\author{Zhong-Qi Ma}
\affiliation {Institute of High Energy Physics, Chinese Academy of
Sciences, Beijing 100049, China}

\author{C. N. Yang}
\email{cnyang@tsinghua.edu.cn} \affiliation {Tsinghua University,
Beijing, China and Chinese University of Hong Kong, Hong Kong}


\begin{abstract}
Conjectures are made for the ground state energy of a large spin 1/2
Fermion system trapped in a 1D harmonic trap with delta function
interaction. States with different spin J are separately studied.
The Thomas-Fermi method is used as an effective test for the
conjecture.
\end{abstract}

\date{\today}
\maketitle


There is recent experimental and theoretical interest \cite{blo,lim}
in a model Hamiltonian for $N$ one dimensional spin $1/2$ Fermions
in a harmonic trap:
\begin{equation}
H=\displaystyle \sum_{i=0}^{N}\left[-\displaystyle \frac{1}{2}
\displaystyle \frac{\partial^{2}}{\partial x_{i}^{2}}+\displaystyle
\frac{1}{2}x_{i}^{2}\right]+g\displaystyle
\sum_{i>j}\delta(x_{i}-x_{j}).
\end{equation}

\noindent For $N=2$, the eigenvalue problem of this Hamiltonian had
been analytically solved \cite{bus}. We concentrate in the present
paper on the ground state energy $E_{J}$ for given total spin $J$ as
$N\rightarrow \infty$. It is well know from group theory that for
spin $J$ the spin wave function is described by the two-row Young
pattern $[N-M,M]$ where $M=(N-2J)/2$ and the space wave function by
its associated pattern \cite{cny}. According to a theorem due to
Lieb and Mattis \cite{lie}, when $g$ is finite,
\[
E_{J}<E_{J'} \qquad {\rm if}~~J<J'.
\]

We notice the following facts:

\noindent (A) As $g\rightarrow +\infty$, the ground state energy
$E_{J}$ for any total spin $J$ approaches \cite{lim} a limit
\begin{equation}
E_{J}\rightarrow \displaystyle \sum_{n=0}^{N-1}\left(\displaystyle
\frac{1}{2} +n\right)=\displaystyle \frac{1}{2}N^{2}. \label{E1}
\end{equation}

\noindent (B) For $g=0$, the ground state wave function for spin
$J=N/2-M$ can be taken to be a product of two determinants
\cite{cny,gir}:
\begin{eqnarray} \Psi &=& \det\left[u_{0}(x_{1})u_{1}(x_{2})\ldots
u_{N-M-1}(x_{N-M})\right]\nonumber \\
&~~~\times& \det\left[u_{0}(x_{N-M+1})\ldots u_{M-1}(x_{N})\right].
\label{psi}
\end{eqnarray}

\noindent where $u_{i}(x)$ is the normalized eigenfunction of the
harmonic oscillator. Its energy is
\begin{equation}
E_{J}=\displaystyle \frac{1}{2}\left[(N-M)^{2}+M^{2}\right].
\label{E2}
\end{equation}

\noindent When $J=0$, $N=2M$ and $E_{0}=N^{2}/4$. Equations
(\ref{E1}) and (\ref{E2}) indicate that
\begin{equation}
1/4 \leq E_{0}/N^{2} \leq 1/2 \qquad {\rm for}~~g\geq 0. \label{E3}
\end{equation}

\noindent (C) For $g\rightarrow -\infty$, when the total spin $J=0$,
$M=N/2$ pairs are formed with spatial size of the order of $-g^{-1}$
each. The internal energy of each pair is $-g^{2}/4$. Thus the $M$
pairs contribute $-Mg^{2}/4$ to the total energy. Between these
pairs there are Fermionic repulsion as well as attractive delta
function interaction. It is difficult to disentangle  this
complicated repulsion-attraction mix. But we observe that the
Fermionic repulsion in absence of attractive $g$ is given by
(\ref{E3}). So it is reasonable to surmise that as $g \rightarrow
-\infty$, the repulsion contributes $X$ to $E_{0}$:
\begin{equation}
E_{0}\rightarrow -\displaystyle \frac{g^{2}}{4}M+X,\qquad {\rm
as}~~g\rightarrow -\infty, \label{E4}
\end{equation}

\noindent where $N^{2}/2\geq X \geq 0$. The $X$ term in (\ref{E4})
is small compared with the other term as $g\rightarrow -\infty$.
Equation (\ref{E4}) holds for $1\leq M \leq N/2$.

Equation (\ref{E3}) shows that for $g\geq 0$, $E_{0}$ is of order
$N^{2}$ while (\ref{E4}) shows that $E_{0}$ is of order $M$ as
$g\rightarrow -\infty$. {\it How can that be?} This question leads
to the following conjecture:

\noindent {\bf \underline{Conjecture 1}}: As $N=2M$ and
$N\rightarrow \infty$, the $E_{0}/N^{2}$ versus $x=g/\sqrt{N}$ curve
approaches a limit:
\begin{equation}
E_{0}/N^{2}\rightarrow f_{0}(g/\sqrt{N}). \label{C1}
\end{equation}

\noindent (\ref{E1}), (\ref{E2}) and (\ref{E4}) show that
\begin{eqnarray}
&&f_{0}(x)\rightarrow 1/2\qquad {\rm as}~~x\rightarrow +\infty,\\
&&f_{0}(0)=1/4,\\
&&f_{0}(x)\rightarrow -x^{2}/8 \qquad {\rm as}~~x\rightarrow
-\infty.
\end{eqnarray}

The schematic figure of conjecture is shown in Fig. 1 and its
generalization is as follows.

\noindent {\bf \underline{Conjecture 2}}: For fixed $J/N=b$, the
$E_{J}/N^{2}$ versus $x=g/\sqrt{N}$ curve approaches a limit
\begin{equation}
E_{J}/N^{2}\rightarrow f_{b}(g/\sqrt{N}), \label{C2}
\end{equation}

\noindent where
\begin{eqnarray}
&&f_{b}(x)\rightarrow 1/2 \qquad {\rm as}~~x\rightarrow +\infty,\\
&&f_{b}=1/4+b^{2},\\
&&f_{b}(x)\rightarrow -\displaystyle \frac{1}{4}\left(\displaystyle
\frac{1}{2}-b\right)x^{2}.
\end{eqnarray}

{\bf \underline{Simple Test of the Two Conjectures}}. It seems
difficult to prove Conjectures 1 and 2. But there is a very simple
but effective test: If Conjectures 1 and 2 are correct, then for
large $N$ and given $J$, the slope at $g=0$ is
\begin{eqnarray}
\displaystyle \frac{dE_{J}}{dg}=N^{3/2}\displaystyle
\frac{d(E_{0}/N^{2})}{d(g/\sqrt{N})}\rightarrow N^{3/2}\displaystyle
\frac{df_{b}(t)}{dt},
\end{eqnarray}

\noindent where $t=g/\sqrt{N}$. Namely, the slope should be
proportional to $N^{3/2}$ for large $N$.

We calculate $dE_{J}/dg$ at $g=0$ by first order perturbation
theory,
\begin{equation}
\displaystyle \frac{dE_{J}}{dg}=\left\langle \left|\displaystyle
\int
\textrm{$\psi$}_{\uparrow}^{\dagger}(x)\textrm{$\psi$}_{\downarrow}^{\dagger}(x)
\textrm{$\psi$}_{\downarrow}(x)\textrm{$\psi$}_{\uparrow}(x)dx\right|\right\rangle
\end{equation}

\noindent where $\textrm{$\psi$}_{\uparrow}$ and
$\textrm{$\psi$}_{\downarrow}$ are fermionic annihilation operators
and the bra \& ket dessignate the ground state at $g=0$. According
to (\ref{psi}) this ground state has $N-M$ spin up particles and $M$
spin down particles in single particle states $u_{0}$, $u_{1}$,
$\ldots$. Thus we have
\[
\displaystyle \frac{dE_{J}}{dg}=\displaystyle
\sum_{i=0}^{N-M-1}\sum_{j=0}^{M-1} \displaystyle \int
u_{i}^{2}(x)u_{j}^{2}(x) dx.
\]

\noindent Defining
\begin{equation}
S[N-M,M]=N^{-3/2}\left.\displaystyle \frac{dE_{J}}{dg}\right|_{g=0},
\end{equation}

\noindent we have
\begin{equation}
S[N-M,M]=N^{-3/2}\displaystyle \int \rho_{N-M}(x)\rho_{M}(x)dx,
\label{S1}
\end{equation}

\noindent where
\begin{equation}
\rho_{M}(x)=\displaystyle \sum_{i=0}^{M-1}u_{i}^{2}(x). \label{rh}
\end{equation}

\noindent The physical meaning of $\rho_{M}(x)$ is the density of
the up-spin (or down-spin) particles at $g=0$.

The density $\rho_{M}(x)$ and the slope $S[N-M,M]$ can be calculated
in a straightforward way with a computer. We define a scaled density
function:
\begin{equation}
R_{M}(y)= \displaystyle \frac{1}{\sqrt{2M}}\rho_{M}(\sqrt{2M} y).
\label{R}
\end{equation}

\noindent From (\ref{rh}) we have, as M approaches infinity,
\begin{equation}
R_{M}(y)\rightarrow 0 \qquad |y|\geq 1. \label{R1}
\end{equation}

\noindent Fig. 2 shows that $R_{M}(y)$ tends to a limit as
$M\rightarrow \infty$ which we shall calculate by the Thomas-Fermi
method later in this paper. Here we list in Table 1 values of the
slope $S[N-M,M]$ for some given parameter $b=J/N$. One can see from
the table that for each value of parameter $b$, $S[N-M,M]$ rapidly
converges to a limit as N increases, {\it supporting both
conjectures 1 and 2}.

\begin{center}

{\bf Table 1 $~~$The slope $S[N-M,M]$ for different $b$. }

\vspace{3mm} {\footnotesize
\begin{tabular}{|c|c||c|c||}
\hline \hline $b=0$&$S[N-M,M]$&$b=1/6$&$S[N-M,M]$\\
\hline
$[1,1]$ & 0.141047 & $[2,1]$ & 0.115165\\
$[3,3]$ & 0.136147& $[4,2]$ & 0.114516\\
$[5,5]$ & 0.135539& $[6,3]$ & 0.114389\\
$[8,8]$ & 0.135292& $[8,4]$ & 0.114343\\
$[10,10]$ & 0.135228& $[10,5]$ & 0.114322\\
$[12,12$ & 0.135191& $[12,6]$ & 0.114311\\
$[15,15]$ & 0.135160& $[14,7]$ & 0.114304\\
$[18,18]$ & 0.135135& $[16,8]$ & 0.114299 \\
$[19,19]$ & 0.135105& $[18,9]$ & 0.114296\\
$[20,20]$ & 0.135097& $[20,10]$ & 0.114294\\ \hline
Eq.(\ref{S3}) & 0.135095 & Eq.(\ref{S3}) & 0.114284 \\
\hline \hline $b=1/4$&$S[N-M,M]$&$b=1/10$&$S[N-M,M]$\\
\hline

$[3,1]$ & 0.0935021& $[3,2]$ & 0.127119\\
$[6,2]$ & 0.0932855& $[6,4]$ & 0.126726\\
$[9,3]$ & 0.0932444& $[9,6]$ & 0.126651\\
$[12,4]$ & 0.0932299& $[12,8]$ & 0.126624\\
$[15,5]$ & 0.0932232& $[15,10]$ & 0.126611\\
$[18,6]$ & 0.0932195& $[18,12]$ & 0.126604\\ \hline
Eq.(\ref{S3}) & 0.0932112 & Eq.(\ref{S3}) & 0.126589 \\
\hline \hline $b=3/10$&$S[N-M,M]$&$b=1/14$&$S[N-M,M]$\\
\hline
$[4,1]$ & 0.0780554& $[4,3]$ & 0.130760\\
$[8,2]$ & 0.0779578& $[8,6]$ & 0.130501\\
$[12,3]$ & 0.0779394& $[12,9]$ & 0.130451\\
$[16,4]$ & 0.0779330& $[16,12]$ & 0.130434 \\
$[20,5]$ & 0.0779300& $[20,15]$ & 0.130429\\ \hline
Eq.(\ref{S3}) & 0.0779247 & Eq.(\ref{S3}) & 0.130411 \\
 \hline \hline

\end{tabular}
}
\end{center}

{\bf \underline{Thomas-Fermi Method}} For large values of $M$, we
can evaluate $\rho_{M}$ by the Thomas-Fermi method (cf. Fig. 3).
When $M$ is very large, the particles in $dx$ have as their maximum
kinetic energy (=maximum total energy minus the potential) $\sim
M-x^{2}/2$. Thus, the number of particles in $dx$ is
$\sqrt{2M-x^{2}}dx/\pi$. I.e.,
\begin{equation}
\rho_{M}(x)\rightarrow \sqrt{2M-x^{2}}/\pi,\qquad {\rm
as}~~M\rightarrow \infty.  \label{rho}
\end{equation}

\noindent Thus from (\ref{R}) we have as $M\rightarrow \infty$
\begin{equation}
R_{M}(y)\rightarrow \sqrt{1-y^{2}}/\pi \qquad {\rm for}~~|y|\lesssim
1. \label{R2}
\end{equation}

Considering (20-23), we obtain the slope $S[N-m,M]$ as $M\rightarrow
\infty$ from (\ref{S1}) \cite{gra}
\begin{eqnarray}
&&\displaystyle \lim_{N\rightarrow \infty}S[N-M,M]=\displaystyle
\frac{2\sqrt{1+2b}}{3\pi}\nonumber \\
&&~~~\times \left[F\left(-\displaystyle \frac{1}{2},\displaystyle
\frac{1}{2};1,z\right) -2bF\left(\displaystyle
\frac{1}{2},\displaystyle \frac{1}{2};1,z\right)\right], \label{S3}
\end{eqnarray}

\noindent where $b=J/N$, $z=(1-2b)/(1+2b)$, and $F$ is the
hypergeometric function. For J=0, (\ref{S3}) reduces to,
\begin{equation}
\displaystyle \lim_{N\rightarrow \infty}S[M,M]=\displaystyle
\frac{4}{3\pi^{2}}\sim 0.135095.
\end{equation}

\noindent The numerical values of (\ref{S3}) are listed in the last
line for each value $b$ in Table 1 for comparison.

\begin{acknowledgments}
One of the author (ZQM) would like to thank Dr. Li-Ming Guan for his
helpful discussion in drawing Figures. This work was partly
supported by the National Natural Science Foundation of China under
Grants No. 10675050.
\end{acknowledgments}

\begin{widetext}

\begin{figure}[htbp]
\includegraphics[scale=.75]
{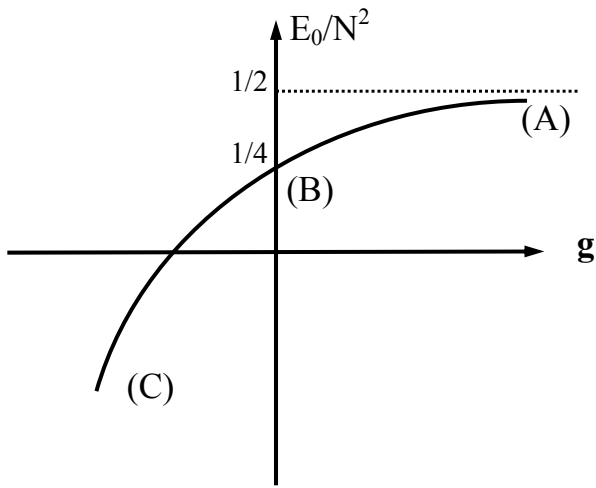}
\caption{Ground state energy $E_J$ vs. $g$ curve for $J=0$ (Schematic). Equations (8) (9) and (10) refer respectively to regions A, B and C in this figure. }
\end{figure}

\begin{figure}[htbp]
\includegraphics[scale=.65]
{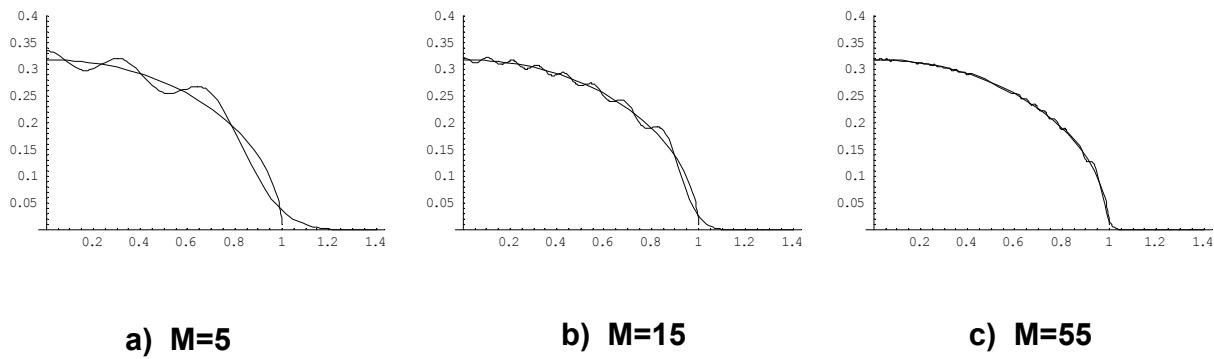}
\caption{Comparison of $R_{M}(y)$ with its limit $\sqrt{1-y^2}/\pi$.}
\end{figure}

\begin{figure}[htbp]
\includegraphics[scale=.75]
{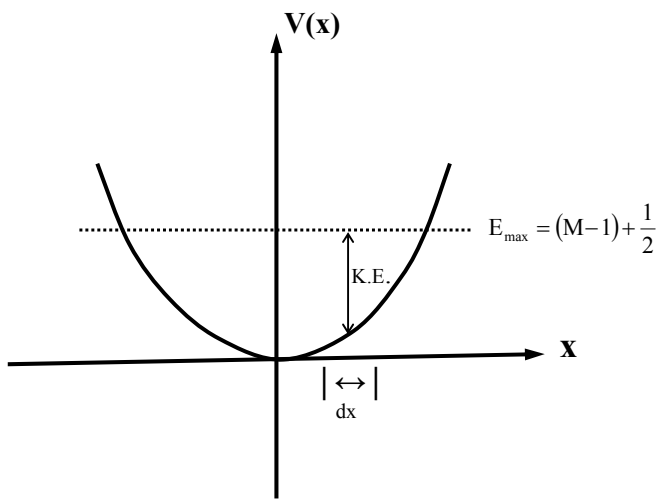}
\caption{Thomas-Fermi Method.}
\end{figure}

\end{widetext}


\begin{thebibliography}{99}

\bibitem{blo}  M. Olshanii, Phys. Rev. Lett. \textbf{81}, 938 (1998);
I. Bloch {\it et al}, Rev. Mod. Phys. \textbf{80}, 885 (2008).

\bibitem{lim} Liming Guan {\it et al}, Phys. Rev. Lett. \textbf{102},
160402 (2009).

\bibitem{bus}  T. Busch {\it et al}, Found. Phys. \textbf{28}, 549
(1998).

\bibitem{cny}  C. N. Yang, \textbf{www.arxiv.org}, 0906.4593 (2009).
We follow the notation of this paper.

\bibitem{lie} E. Lieb and D. Mattis, Phys. Rev. \textbf{125},
164 (1962).

\bibitem{gir}  M. D. Girardeau and A. Minguzzi, Phys. Rev. Lett.
\textbf{99}, 230402 (2007).

\bibitem{gra} I. S. Gradshteyn and I. M. Ryzhik, {\it Table of Integrals,
Series, and Products}, (Academic Press, New York, 1980).


\end{thebibliography}
\end{document}